\documentclass[a4paper,11pt]{article}

\usepackage{pos}
\usepackage{orcidlink}
\usepackage{amsmath}
\usepackage{booktabs}
\usepackage[small]{subfigure}

\newcommand{\nn}{\nonumber}
\newcommand{\as}{\alpha_s}
\newcommand{\mathd}{\mathrm{d}}
\newcommand{\Gamt}{\Gamma_{\!t}}
\newcommand{\LQCD}{\Lambda_\mathrm{QCD}}
\newcommand{\bare}{\mathrm{bare}}
\newcommand{\rc}{\mathrm{c}}
\newcommand{\rd}{\mathrm{d}}
\newcommand{\ord}{\mathcal{O}}
\newcommand{\bra}[1]{\left\langle #1 \right|}
\newcommand{\ket}[1]{\left| #1 \right\rangle}
\newcommand{\cB}{\mathcal{B}}
\newcommand{\MSb}{\overline{\mathrm{MS}}}

\DeclareRobustCommand{\fig}[1]{Fig.~\ref{fig:#1}}
\DeclareRobustCommand{\tab}[1]{Table~\ref{tab:#1}}
\DeclareRobustCommand{\eqsss}[3]{Eqs.~\eqref{eq:#1}, \eqref{eq:#2} and \eqref{eq:#3}}

\title{Three-loop jet function for boosted top quarks}

\author*[a]{Vicent Mateu}
\author[a]{Alberto M.~Clavero}
\author[b]{Maximilian Stahlhofen}

\affiliation[a]{Departamento de F\'isica Fundamental e IUFFyM,\\Universidad de Salamanca, E-37008 Salamanca, Spain}

\affiliation[b]{Albert-Ludwigs-Universit\"at Freiburg, Physikalisches Institut,\\D-79104 Freiburg, Germany}

\emailAdd{vmateu@usal.es}
\emailAdd{albertomarcla@usal.es}
\emailAdd{maximilian.stahlhofen@physik.uni-freiburg.de}

\abstract{We present the calculation of the inclusive jet function for highly energetic heavy quarks at order $\mathcal{O}(\alpha_s^3)$ using boosted Heavy-Quark Effective Theory (bHQET). This jet function describes the effect of collinear radiation emitted by energetic heavy quarks on observables dependent on the jet invariant mass $M$. In particular, we focus on the regime $M^2 - m^2 \ll m^2$, which is relevant for boosted top quark production at high-energy colliders in the resonance region. Our results are consistent with non-Abelian exponentiation and reproduce the known cusp and non-cusp anomalous dimensions up to three loops. We also verify that the $n_\ell^2 \alpha_s^3$ contribution, with $n_\ell$ denoting the number of light quark flavors, agrees with predictions from renormalon calculus. This calculation completes the list of ingredients required for the N$^3$LL$^\prime$ resummed (self-normalized) thrust distribution, an essential component for calibrating the top quark mass parameter in parton-shower Monte Carlo generators. It likewise contributes to the invariant-mass distribution of reconstructed top quarks, enabling precise mass determinations at future lepton colliders. Finally, we determine the relation between the pole and two short-distance jet-mass schemes at $\mathcal{O}(\alpha_s^3)$ and provide an estimate of the non-logarithmic part of the four-loop jet function based on renormalon dominance.}

\FullConference{The European Physical Society Conference on High Energy Physics (EPS-HEP2025)\\
7-11 July 2025\\
Marseille, France\\}

\begin{document}
\maketitle

\section{Introduction}

Precise determinations of the top quark mass play a central role in testing the Standard Model (SM). Its value critically impacts the electroweak vacuum stability and affects other SM parameters such as the $W$- and Higgs-boson masses. Obtaining a precise measurement of the top mass $m$ is therefore essential for performing consistency checks of the SM.

Since the top quark decays before it can hadronize, many analyses attempt to determine its mass using kinematic information from its decay products. These direct experimental measurements have reached a precision of approximately $300\,$MeV at the Tevatron and LHC, and prospects indicate that this uncertainty can be further reduced.
The mass determined by direct measurements does not, however, correspond to any well-defined mass within quantum field theory, but rather to a mass parameter inherent to parton-shower Monte Carlo (MC) generators, dubbed $m^{\rm MC}$.

Event shapes at $e^+e^-$ colliders have been extensively used to probe the QCD structure, tune MC generators, and determine the strong coupling $\as$ with high precision. For processes involving boosted top quark pairs, it has been shown that the top mass can be determined (in a suitable short-distance scheme) with uncertainty smaller than $\LQCD$ using event shapes related to hemisphere invariant masses \cite{Fleming:2007qr,Fleming:2007xt}.
The maximal sensitivity to the top mass occurs in the peak region of the hemisphere mass
distribution, characterized by $M^2 - m^2\equiv m \hat{s}$ where $m \gg \hat s \sim \Gamt \gg \LQCD$. In this region, highly energetic top quarks can be described using boosted Heavy Quark Effective Theory (bHQET) matched to Soft Collinear Effective Theory (SCET). This EFT framework enables the resummation of large logarithms, consistently includes the top decay width $\Gamt$ and accounts for soft hadronization power corrections from first
principles.

\vspace{0.5em}

At leading power in the expansion in powers of $m/Q$, $\Gamt/m$, and $\hat{s}/m$ (with $Q$ the center-of-mass energy), the dijet factorization theorem for the double hemisphere invariant mass differential distribution reads~\cite{Fleming:2007qr}
\begin{align}
\frac{\mathd^2 \sigma^{(\mathrm{dijet})}}{\mathd M_1^2 \mathd M_{2}^2} ={}& \sigma_0\, H^{(n_\ell+1)}_Q\!\bigl(Q, \mu\bigr)\, H_m\biggl(m, \frac{Q}{m}, \mu\biggr)\!
\int \! \mathd \ell^{+} \mathd \ell^{-} \,
S^{(n_\ell)}\bigl(\ell^{+}, \ell^{-}, \mu\bigr) \\
&\times
B^{(n_\ell)}_n\!\biggl(\frac{M_1^2-Q \ell^{+}}{m}-m, \Gamt, \mu \biggr)\,
B^{(n_\ell)}_{\bar n}\!\biggl(\frac{M_2^2-Q \ell^{-}}{m}-m, \Gamt, \mu \biggr).\nn
\end{align}

This formula contains the massless hard $H_Q$ and soft functions $S$ coming from the corresponding SCET factorization theorem, which are known at three and two loops, respectively.
The matching factor $H_m$ encodes the effect of soft and collinear mass-modes, and its rapidity-resummed expression is currently known at NNLL$^\prime$ (N$^3$LO)~\cite{Fael:2022miw}. The bHQET jet functions $B_n$ and $B_{\bar n}$, identical due to charge conjugation, describe the dynamics of the back-to-back boosted heavy-quark jets. The main aim of our work is to extend the calculation of the jet function $B_n$, previously determined up to two-loop order \cite{Jain:2008gb}, to three loops. Details of this calculation have been published in Ref.~\cite{Clavero:2024yav}.

Our result can be used to obtain the di-hemisphere mass, thrust, heavy jet mass and \mbox{C-parameter} cross sections in the peak region at N$^3$LL$^\prime$.
In addition, it is the last missing piece to achieve an N$^3$LL$^\prime$-precise calibration between $m^{\rm MC}$ and the MSR mass $m^{\rm MSR}$~\cite{Hoang:2017suc}.
Finally, it will eventually improve the precision of (indirect) top mass determinations from boosted-top production at a future lepton collider~\cite{Butenschoen:2016lpz,Dehnadi:2023msm}.

\section{Theoretical setup}

\begin{figure*}[t]
\centering
\subfigure{\includegraphics[width=0.3\textwidth]{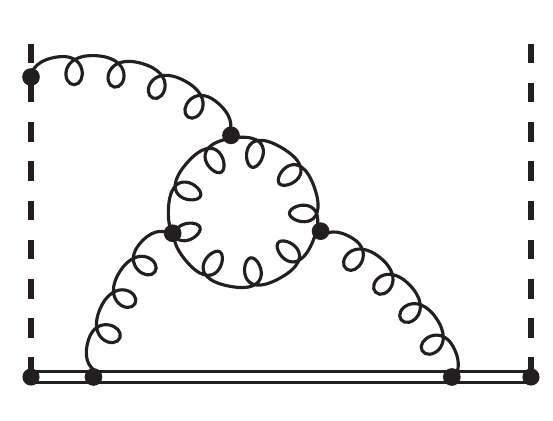}}
\subfigure{\includegraphics[width=0.3\textwidth]{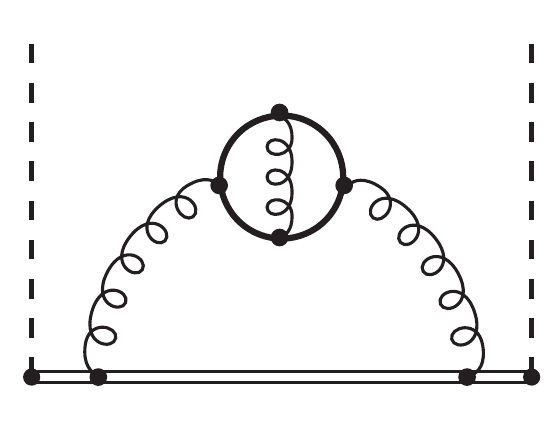}}
\subfigure{\includegraphics[width=0.3\textwidth]{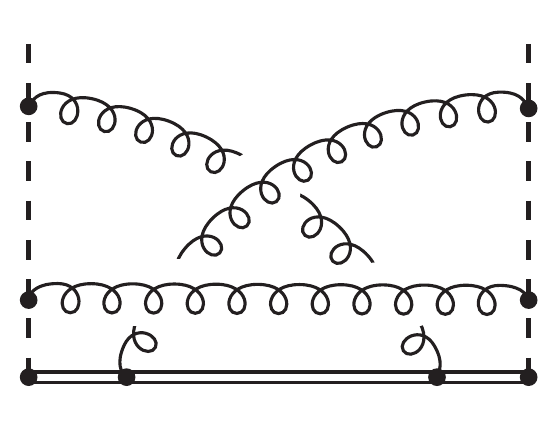}}
\caption{Sample of Feynman diagrams contributing to the jet function at three loops. Double, curly, dashed and solid lines represent heavy-quarks, gluons, light-like Wilson lines and light quarks, respectively.}%
\label{fig:FDEx}%
\end{figure*}

The bHQET jet function for stable top quarks with heavy-quark mass expressed in the pole scheme is defined through the forward-scattering matrix element \cite{Fleming:2007qr,Jain:2008gb}%
\begin{equation}
\mathcal{B}^\bare (\hat{s}) \equiv\! \frac{-
i}{4 \pi N_c m} \!\int\! \rd^d z\, e^{i r \cdot z} \!\bra{ \,0\, }
T \bigl\{ \bar{h}_{v} (0) W_n (0) W_n^{\dag} (z) h_{v} (z) \bigr\}\! \ket{\,0\,}\!
= - \frac{1}{\pi m \hat{s}} \,+ \ord(\as)\,,
\label{eq:def_Bbare}
\end{equation}
where $h_v$ is the bHQET heavy-quark field with velocity $v^\mu$, such that $\hat s = 2 v \cdot r$ and $v^2=1$. $W_n$ is the $n$-collinear Wilson line, with $n^\mu=(1,\vec{n})$ a light-like vector ($|\vec{n}|=1$) and $\vec{n}$ collinear to the jet direction. Finally, $N_c$ denotes the number of colors.
The renormalized bHQET jet function $B \equiv B_n$ is obtained through the imaginary part of the matrix element in Eq.~\eqref{eq:def_Bbare},
\begin{equation}
B (\hat s) \equiv \mathrm{Im} \bigl[ \mathcal{B} (\hat{s} + i \eta) \bigr] = \frac{1}{m} \delta(\hat{s}) + \ord(\as),
\label{eq:def_B}
\end{equation}
where $\eta$ is a positive infinitesimal. The expression for unstable quarks is obtained by shifting $\eta \to \Gamt$ in Eq.~\eqref{eq:def_B}.
In order to efficiently solve the jet function's renormalization group equation (RGE), it is convenient to work with the Fourier transform of $B$, also referred to as the position-space jet function, and its exponentiated form

\begin{equation}
\tilde{B}(x)\equiv \!\int \rd \hat s\,e^{-i (x - i \eta) \hat s} B(\hat s)\,,\qquad
\tilde B(x)= \frac{1}{m} \!\exp\! \big[ \tilde{b}(x) \big]\,.
\label{eq:def_Bx}
\end{equation}
As argued in Ref.~\cite{Jain:2008gb}, the jet function is subject to non-Abelian exponentiation, which means that the exponent $\tilde{b}(x)$ exclusively contains fully connected color factors~\cite{Gardi:2013ita}. This class of color factors corresponds to that of Feynman diagrams that remain connected once all Wilson lines are removed from the graph. In particular, this implies that, at $\ord(\as^3)$, the color structures $C_F^2C_A$ and $C_F^3$ must be absent and only diagrams involving a fermion loop with four gluon attachments, like the second diagram in \fig{FDEx}, can contribute to the $C_F^2T_Fn_\ell$ term of $\tilde{b}(x)$.

\vspace{0.5em}

To regulate UV and IR divergences in loop diagrams, we employ dimensional regularization with $d = 4-2\varepsilon$. For the renormalization of the jet function and the strong coupling we
employ the $\MSb$ scheme, while the heavy-quark mass $m$ is defined (for the time being) in the pole scheme.
The divergences in the momentum-space bare jet function are removed by convolving it with a renormalization factor $Z_B (\hat s, \mu)$ which depends on the $\MSb$ renormalization scale $\mu$.

Hence, the renormalized jet function is $\mu$-dependent and satisfies a RGE, which in position space reads
\begin{equation}
\dfrac{\rd \log \bigl[m\tilde{B}(x,\mu)\bigr]}{\rd \log \mu}=
2\Gamma^\rc[\as(\mu)] \log (i e^{\gamma_E} x \mu) + \gamma_B^\text{nc}[\as(\mu)]\,.
\end{equation}
$\Gamma^c$ and $\gamma_B^{\rm nc}$ are the cusp and non-cusp anomalous dimensions, respectively. The solution of the RGE in momentum space enables resummation of large logarithms of the type $\log (\hat{s}/m)$ through scale evolution. In our paper \cite{Clavero:2024yav} we provide relations among the coefficients appearing in the perturbative expansion of the various versions of the jet function, like the ones in \eqsss{def_Bbare}{def_B}{def_Bx}. From the RGE, we also derive relations that express the logarithmic coefficients appearing in the renormalized jet function in terms of the cusp, non-cusp and strong-coupling anomalous dimensions and lower-loop non-logarithmic coefficients. These relations enable efficient code implementation of the RGE and conversion between jet function variants.

\vspace{0.5em}

The jet function as defined in Eq.~\eqref{eq:def_B} exhibits an infrared renormalon ambiguity. The renormalon leads to an asymptotically divergent behavior of the perturbative series in $\as$ unless it is canceled by switching to a low-scale short-distance scheme such as the MSR mass. The corresponding scheme change is implemented by shifting the argument of the jet function,
\begin{equation}
B(\hat s, \delta m, \Gamt, \mu) = \text{Im} \bigl[ \mathcal{B}(\hat{s} -2 \delta m + i \Gamt, \mu) \bigr]
=\exp\biggl(\!-2\delta m\frac{\partial}{\partial \hat s}\biggr)B(\hat s, 0,\Gamt, \mu) \,,
\end{equation}
where $\delta m = m - m_{\rm SD}$ defines the short-distance mass $m_{SD}$. We study two alternative jet-mass schemes: the \textit{derivative scheme} \cite{Jain:2008gb} and the \textit{non-derivative scheme} \cite{Gracia:2021nut}, defined respectively through
\begin{align}\label{eq:short-dist}
\delta m^J(\mu,R)
={}&
\frac{Re^{\gamma_E}}{2} \biggl\{\frac{\mathd}{\mathd \log(ix)}\log\bigl[ m \tilde{B}(x, \mu)\bigr] \biggr\}_{i x e^{\gamma_E}=1 / R}\, , \\
\delta m^{J^\prime}\!(R) ={}& \frac{Re^{\gamma_E}}{2} \log\Bigl[ m \tilde{B}\Bigl(\frac{1}{i R e^{\gamma_E}}, R\Bigr)\Bigr]\, .\nonumber
\end{align}
Both schemes remove the renormalon and exhibit different perturbative behavior, which we compare to the MSR mass as a benchmark. Since $\delta m ^{J^\prime}$ does not involve any derivatives, the asymptotic behavior sets in faster than for the original jet-mass scheme, which may be advantageous in situations where only few perturbative orders are known. On the other hand, the derivative jet-mass scheme requires less perturbative ingredients at a given finite order than the non-derivative scheme \cite{Clavero:2024yav}.

\begin{table}[t]
\centering
\begin{tabular}{cccc}
\toprule
& $1$ loop & $2$ loops & $3$ loops \\ \midrule
Feynman diagrams & $4$ & ${\sim}50$ & ${\sim}1100$ \\
Scalar integrals & $3$ & ${\sim}70$ & ${\sim}5400$ \\
Master integrals & $1$ & $3$ & $20$ \\
\bottomrule
\end{tabular}
\caption{Number of Feynman diagrams, non-vanishing scalar integrals, and master integrals contributing to the jet function at one, two and three loops. Numbers may vary depending on the particular setup, and should be understood as just a rough complexity measure.} \label{tab:estimate_integrals}
\end{table}

\section{Computation of the jet function}

Our computation of the bare jet function matrix element in Eq.~\eqref{eq:def_Bbare}, which in many aspects resembles that of the soft function for heavy-to-light quark decays~\cite{Bruser:2019yjk}, proceeds as follows:
\begin{figure*}[t]
\subfigure
{\includegraphics[width=0.45\textwidth]{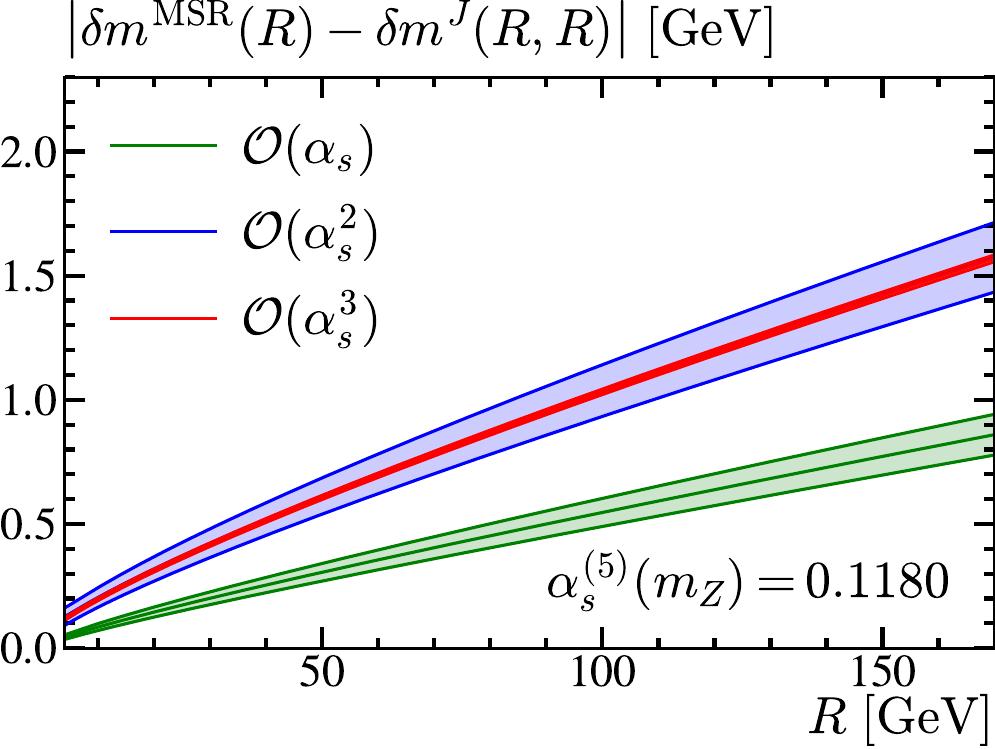}\vspace*{-1cm}}
~~ \subfigure
{\includegraphics[width=0.45\textwidth]{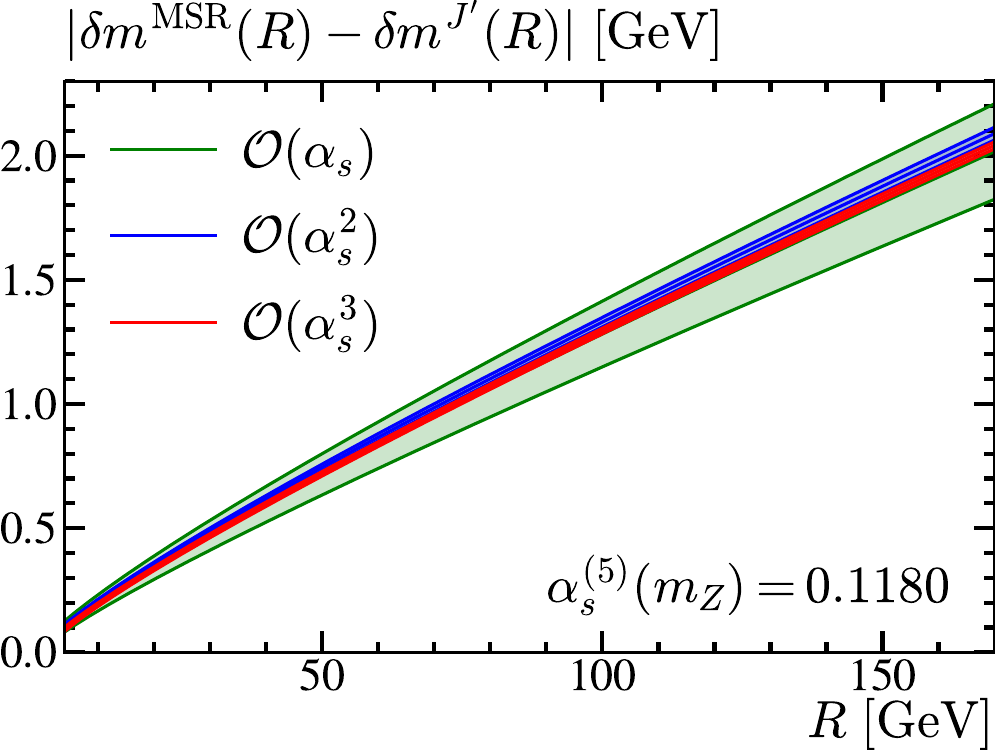}\vspace*{-1cm}}
\caption{Difference between the heavy quark MSR and jet masses in the derivative~(left) and non-derivative~(right) schemes in absolute value. In the plots we use $n_\ell=5$ for the number of active light flavors and $\alpha_s^{(n_\ell=5)}(m_Z)=0.1180$. For both masses we expand $\alpha_s(R)$ in terms of $\alpha_s(\mu)$, and generate uncertainty bands varying $\mu$ in the range $R/2$ to $2R$. \label{fig:masses}}
\end{figure*}
\begin{enumerate}
\item Generation of Feynman diagrams contributing to the matrix element using \texttt{qgraf}~\cite{Nogueira:1991ex}.
\item Mapping of the diagrams to integral families based on the denominator structure. For this procedure we use the algorithm outlined in Ref.~\cite{Pak:2011xt} implemented in \texttt{Looping} \cite{Looping}.
\item Reduction to scalar integrals using \texttt{FORM}~\cite{Ruijl:2017dtg} and the \texttt{color.h} package \cite{vanRitbergen:1998pn}.
Linear dependencies between denominators are removed by applying the multivariate partial fractioning algorithm of Ref.~\cite{Pak:2011xt}. Scaleless integrals are discarded and symmetries related to momentum shifts are exploited to reduce the number of occurring integrals. All these operations are automated by the \texttt{Looping} interface.
\item IBP reduction, using \texttt{FIRE} \cite{Smirnov:2019qkx} and \texttt{LiteRed} \cite{Lee:2013mka}, to express all scalar integrals in terms of a basis of master integrals (MIs). Additional relations between MIs are found through dimensional recurrence relations following~\cite{Bruser:2018rad}.
\item Analytical calculation of the MIs. We employed the method described in Ref.~\cite{vonManteuffel:2014qoa}, which relies on a basis of (quasi-)finite integrals that can be evaluated with \texttt{HyperInt}~\cite{Panzer:2014caa}, and Mellin-Barnes techniques.
\item All diagrams are added up and the results for the MIs are inserted to obtain the bare jet function matrix element $\cB^\bare(\hat{s})$. Taking the imaginary part yields the bare jet function.
\end{enumerate}

To illustrate the variety of three-loop Feynman diagrams involved in the computation, we show a selection of them in \fig{FDEx}. The number of Feynman diagrams, non-vanishing scalar integrals, and MIs relevant for our jet function calculation at one, two, and three loops (which serves as a rough complexity measure) is displayed in \tab{estimate_integrals}.

\section{Main results}
Our main result is the non-logarithmic term of the exponent of the renormalized position-space bHQET jet function
\begin{equation}\label{eq:exponent}
\tilde b(x) = \sum^\infty_{l=1} \Big(\dfrac{\as}{4 \pi}\Big)^l \sum^{l+1}_{k=0}\tilde b_{lk}\, \log^k (i e^{\gamma_E} x \mu).
\end{equation}
at three loops (for $N_c=3$),
\begin{align}
\tilde b_{30}
={}&
\biggl( \frac{203 \pi ^2 \zeta_3}{27}-\frac{105398 \zeta_3}{243}+\frac{236 \zeta_3^2}{9} + \frac{902 \zeta_5}{9} + \frac{31952191}{26244} + \frac{93821 \pi
^2}{8748}
\label{eq:JF_RESULT}
\\ \nonumber & -\frac{3023 \pi ^4}{4860} + \frac{1031 \pi ^6}{10206} \biggr) C_FC_{A}^2
+
\biggl( \frac{3488 \zeta_3}{243}+\frac{846784}{6561}-\frac{8 \pi ^2}{243}+\frac{52 \pi ^4}{1215} \biggr) C_FT^2_F n_\ell ^2
\\ \nonumber
& + \biggl( \frac{10760 \zeta_3}{81}+\frac{8 \pi ^2 \zeta_3}{9}+\frac{224 \zeta_5}{9}-\frac{124717}{486}-\frac{55 \pi ^2}{54}+\frac{148 \pi ^4}{405}\biggr) C_F^2 T_F n_\ell\\ \nonumber
& + \biggl( \frac{1664 \zeta_3}{81}-\frac{76 \pi ^2 \zeta_3}{27}-\frac{88 \zeta_5}{3}-\frac{5273287}{6561}-\frac{12793 \pi
^2}{2187}-\frac{421 \pi ^4}{1215} \biggr) C_FC_A T_F n_\ell\, \\
={}& 50.054 n^2_\ell - 1899.8 n_\ell + 12834\, \nonumber,
\end{align}
where $\tilde b_{30}$ is defined through the expansion series of the exponent of the position-space jet function in Eq.~\eqref{eq:exponent}.
The logarithmic terms of the exponent can be derived from the known anomalous dimensions and beta function coefficients in the $\as/(4\pi)$ expansion,
\begin{equation}\label{eq:nonRec}
\tilde{b}_{l 1} = \gamma^B_{l-1} + 2\!\sum_{j = 1}^{l -1} j \beta_{l - j - 1}\, \tilde{b}_{j0}\,,\quad
\tilde{b}_{l 2} =\Gamma^\rc_{\!l-1} + \sum_{j = 1}^{l -1} j \beta_{l - j - 1}\, \tilde{b}_{j1}\,,
\quad
\tilde{b}_{lk} = \frac{2}{k} \sum_{j =k - 2}^{l -1} \!\!\!j \beta_{l - j - 1} \tilde{b}_{j, k - 1}\,.
\end{equation}
We also recomputed the one- and two-loop bare matrix element for arbitrary dimension $d$. Results can be found in our paper \cite{Clavero:2024yav}.
We validated our results through multiple stringent checks. We have performed the computation of the bare matrix element in a general covariant gauge and checked that the dependence on the gauge parameter drops out once expressed in terms of MIs. Our expression for the logarithm of the position-space jet function in Eq.~\eqref{eq:JF_RESULT} is consistent with non-Abelian exponentiation. Furthermore, we have checked that $\cB^\bare$ is free of $C_F^3$ and $C_F^2C_A$, and only diagrams involving a fermion loop with four gluon attachments can contribute to the $C_F^2 T_F n_\ell$ term already before the $\varepsilon = (4-d)/2$ expansion. We also verified that the $C_FT_F^2 n_\ell^2$ term in Eq.~\eqref{eq:JF_RESULT} agrees with the prediction from renormalon calculus in Ref.~\cite{Gracia:2021nut}, and our full computation reproduces the known three-loop anomalous dimension coefficients $\gamma_2^B$ and $\Gamma_2^c$~\cite{Korchemsky:1987wg, Moch:2004pa,Hoang:2015vua,Bruser:2019yjk,Becher:2008cf}. Additionally, all MIs were checked numerically using \texttt{pySecDec}~\cite{Heinrich:2023til} and \texttt{FIESTA}~\cite{Smirnov:2021rhf} to $\ord(\varepsilon^3)$.

\subsection{Jet-Mass schemes and four-loop jet function estimate}

Using our three-loop result, we obtain the relation between the pole and jet-mass schemes defined in Eq.~\eqref{eq:short-dist} at
$\ord(\as^3)$. The explicit expressions are lengthy and provided in our paper. Figure.~\ref{fig:masses} compares the two jet-mass schemes to the MSR mass benchmark.
Our analysis shows that the non-derivative scheme exhibits faster convergence, with the
one-loop result already falling within the two- and three-loop uncertainty bands. The
derivative scheme requires at least two loops for reliable results. Both schemes effectively
remove the pole-mass renormalon at higher orders.

Based on renormalon dominance, we estimate the four-loop non-logarithmic
coefficient using the R-evolution formalism applied to the non-derivative jet-mass scheme. We refer to \cite{Gracia:2021nut} for details of the procedure.
Specifically, we predict $\tilde b_{40}(n_\ell) = \sum_{i=0}^3 \tilde b_4^i \, n_\ell^i\,$, the non-logarithmic four-loop coefficient of the position-space jet function's exponent for different numbers of light flavors $n_\ell$.

\begin{figure*}[t]
\subfigure
{\includegraphics[width=0.495\textwidth]{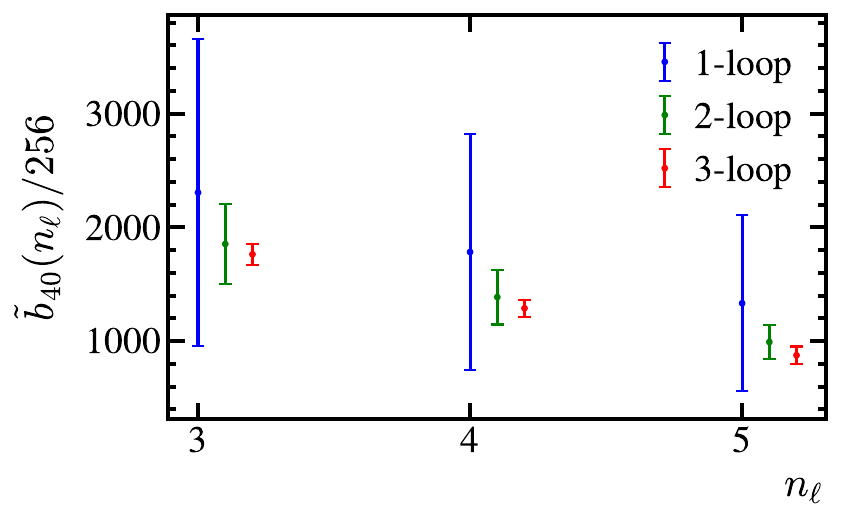}%
\label{fig:4loop}}
\subfigure
{\includegraphics[width=0.495\textwidth]{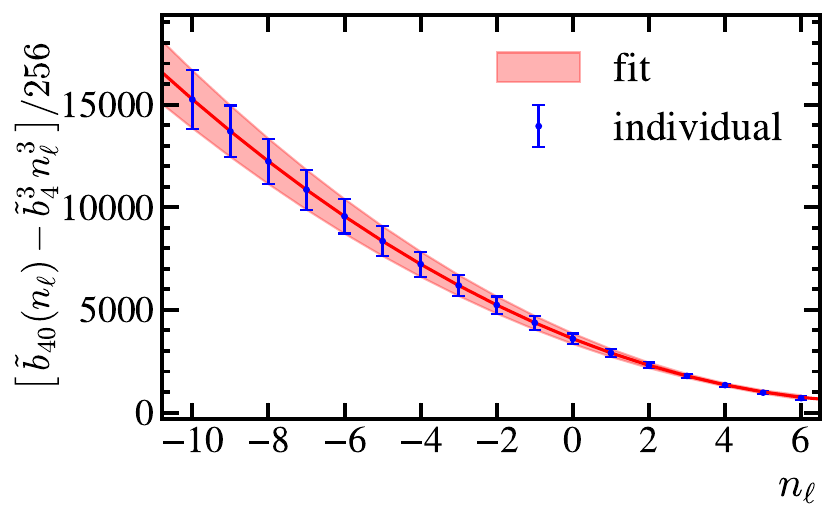}%
\label{fig:fit}}
\caption{Estimates of $\tilde{b}_{40}$ for various numbers of light flavors $n_\ell$ with $N_c=3$. In the left panel, we show the estimates when including lower-order information up to one (blue), two (green), and three (red) loops. In the right panel, we use three-loop input to make predictions for a large set of $n_\ell$ values (in blue), and show as a red band the predictions obtained using the results for the various flavor coefficients. For details in the uncertainty calculation, we refer to our paper Ref.~\cite{Clavero:2024yav}.\label{fig:b4}}
\end{figure*}

Our results are shown in \fig{b4}. In the left panel, we present estimates of $\tilde b_{40}$ for various $n_\ell$ values, utilizing lower-order information up to one (blue), two (green), and three (red) loops. In the right panel, we use three-loop input to make predictions for a wide range of $n_\ell$ values (in blue). The red band represents predictions obtained using the results for the various flavor coefficients, along with the corresponding uncertainties generated through error propagation. Our findings indicate that incorporating higher-order information leads to more precise estimates of $\tilde b_{40}$.

\section{Conclusions}

We have computed the inclusive bHQET jet function for boosted heavy quarks to three-loop
accuracy, providing a universal ingredient in peak-region factorization theorems
for event shape observables probing the top mass. Our result represents the last missing piece for N$^3$LL$^\prime$ resummed predictions of the self-normalized thrust distributions used in Monte
Carlo top mass calibration. We have also derived universal relations between different representations of the jet function
and compared suitable short-distance mass schemes.

Future directions include computing the four-loop non-cusp anomalous dimension to achieve N$^4$LL accuracy on the self-normalized thrust prediction, including effects of massive bottom quarks in the loop corrections, computing the missing three-loop contributions to $H_m$ for full N$^3$LL$^\prime$ predictions, and updating the calibration of the Monte-Carlo top mass to N$^3$LL$^\prime$ accuracy using our new result~\cite{Butenschoen:2016lpz,Dehnadi:2023msm}.

\begin{acknowledgments}
This work was supported in part by the Spanish MICIU/AEI/10.13039/501100011033 grant No.~PID2022-141910NB-I00 and the JCyL grant SA091P24 under program EDU/841/2024. A.\,M.\,C.\ is supported by a FPU scholarship funded by the Spanish MICIU under grant no.\ FPU22/02506.
\end{acknowledgments}

\bibliography{BHQET}

\providecommand{\href}[2]{#2}\begingroup\raggedright\begin{thebibliography}{10}

\bibitem{Fleming:2007qr}
S.~Fleming, A.~H. Hoang, S.~Mantry and I.~W. Stewart, \emph{{Jets from massive unstable particles: Top-mass determination}}, \href{http://dx.doi.org/10.1103/PhysRevD.77.074010}{\emph{Phys. Rev.} {\bfseries D77} (2008) 074010}, [\href{https://arxiv.org/abs/hep-ph/0703207}{{\ttfamily hep-ph/0703207}}].

\bibitem{Fleming:2007xt}
S.~Fleming, A.~H. Hoang, S.~Mantry and I.~W. Stewart, \emph{{Top Jets in the Peak Region: Factorization Analysis with NLL Resummation}}, \href{http://dx.doi.org/10.1103/PhysRevD.77.114003}{\emph{Phys. Rev.} {\bfseries D77} (2008) 114003}, [\href{https://arxiv.org/abs/0711.2079}{{\ttfamily 0711.2079}}].

\bibitem{Fael:2022miw}
M.~Fael, F.~Lange, K.~Sch\"onwald and M.~Steinhauser, \emph{{Singlet and nonsinglet three-loop massive form factors}}, \href{http://dx.doi.org/10.1103/PhysRevD.106.034029}{\emph{Phys. Rev. D} {\bfseries 106} (2022) 034029}, [\href{https://arxiv.org/abs/2207.00027}{{\ttfamily 2207.00027}}].

\bibitem{Jain:2008gb}
A.~Jain, I.~Scimemi and I.~W. Stewart, \emph{{Two-loop Jet-Function and Jet-Mass for Top Quarks}}, \href{http://dx.doi.org/10.1103/PhysRevD.77.094008}{\emph{Phys. Rev.} {\bfseries D77} (2008) 094008}, [\href{https://arxiv.org/abs/0801.0743}{{\ttfamily 0801.0743}}].

\bibitem{Clavero:2024yav}
A.~M. Clavero, R.~Br{\"u}ser, V.~Mateu and M.~Stahlhofen, \emph{{Three-loop jet function for boosted heavy quarks}}, \href{http://dx.doi.org/10.1007/JHEP04(2025)040}{\emph{JHEP} {\bfseries 04} (2025) 040}, [\href{https://arxiv.org/abs/2412.06881}{{\ttfamily 2412.06881}}].

\bibitem{Hoang:2017suc}
A.~H. Hoang, A.~Jain, C.~Lepenik, V.~Mateu, M.~Preisser, I.~Scimemi et~al., \emph{{The MSR mass and the $\mathcal{O}\left({\Lambda}_{\mathrm{QCD}}\right)$ renormalon sum rule}}, \href{http://dx.doi.org/10.1007/JHEP04(2018)003}{\emph{JHEP} {\bfseries 04} (2018) 003}, [\href{https://arxiv.org/abs/1704.01580}{{\ttfamily 1704.01580}}].

\bibitem{Butenschoen:2016lpz}
M.~Butenschoen, B.~Dehnadi, A.~H. Hoang, V.~Mateu, M.~Preisser and I.~W. Stewart, \emph{{Top Quark Mass Calibration for Monte Carlo Event Generators}}, \href{http://dx.doi.org/10.1103/PhysRevLett.117.232001}{\emph{Phys. Rev. Lett.} {\bfseries 117} (2016) 232001}, [\href{https://arxiv.org/abs/1608.01318}{{\ttfamily 1608.01318}}].

\bibitem{Dehnadi:2023msm}
B.~Dehnadi, A.~H. Hoang, O.~L. Jin and V.~Mateu, \emph{{Top quark mass calibration for Monte Carlo event generators \textemdash{} an update}}, \href{http://dx.doi.org/10.1007/JHEP12(2023)065}{\emph{JHEP} {\bfseries 12} (2023) 065}, [\href{https://arxiv.org/abs/2309.00547}{{\ttfamily 2309.00547}}].

\bibitem{Gardi:2013ita}
E.~Gardi, J.~M. Smillie and C.~D. White, \emph{{The Non-Abelian Exponentiation theorem for multiple Wilson lines}}, \href{http://dx.doi.org/10.1007/JHEP06(2013)088}{\emph{JHEP} {\bfseries 06} (2013) 088}, [\href{https://arxiv.org/abs/1304.7040}{{\ttfamily 1304.7040}}].

\bibitem{Gracia:2021nut}
N.~G. Gracia and V.~Mateu, \emph{{Toward massless and massive event shapes in the large-\ensuremath{\beta}$_{0}$ limit}}, \href{http://dx.doi.org/10.1007/JHEP07(2021)229}{\emph{JHEP} {\bfseries 07} (2021) 229}, [\href{https://arxiv.org/abs/2104.13942}{{\ttfamily 2104.13942}}].

\bibitem{Bruser:2019yjk}
R.~Br\"user, Z.~L. Liu and M.~Stahlhofen, \emph{{Three-loop soft function for heavy-to-light quark decays}}, \href{http://dx.doi.org/10.1007/JHEP03(2020)071}{\emph{JHEP} {\bfseries 03} (2020) 071}, [\href{https://arxiv.org/abs/1911.04494}{{\ttfamily 1911.04494}}].

\bibitem{Nogueira:1991ex}
P.~Nogueira, \emph{{Automatic Feynman Graph Generation}}, \href{http://dx.doi.org/10.1006/jcph.1993.1074}{\emph{J. Comput. Phys.} {\bfseries 105} (1993) 279--289}.

\bibitem{Pak:2011xt}
A.~Pak, \emph{{The Toolbox of modern multi-loop calculations: novel analytic and semi-analytic techniques}}, \href{http://dx.doi.org/10.1088/1742-6596/368/1/012049}{\emph{J. Phys. Conf. Ser.} {\bfseries 368} (2012) 012049}, [\href{https://arxiv.org/abs/1111.0868}{{\ttfamily 1111.0868}}].

\bibitem{Looping}
R.~Br{\"u}ser, \emph{{Looping}}, {\emph{unpublished\!} }.

\bibitem{Ruijl:2017dtg}
B.~Ruijl, T.~Ueda and J.~Vermaseren, \emph{{FORM version 4.2}},  \href{https://arxiv.org/abs/1707.06453}{{\ttfamily 1707.06453}}.

\bibitem{vanRitbergen:1998pn}
T.~van Ritbergen, A.~N. Schellekens and J.~A.~M. Vermaseren, \emph{{Group theory factors for Feynman diagrams}}, \href{http://dx.doi.org/10.1142/S0217751X99000038}{\emph{Int. J. Mod. Phys. A} {\bfseries 14} (1999) 41--96}, [\href{https://arxiv.org/abs/hep-ph/9802376}{{\ttfamily hep-ph/9802376}}].

\bibitem{Smirnov:2019qkx}
A.~V. Smirnov and F.~S. Chuharev, \emph{{FIRE6: Feynman Integral REduction with Modular Arithmetic}}, \href{http://dx.doi.org/10.1016/j.cpc.2019.106877}{\emph{Comput. Phys. Commun.} {\bfseries 247} (2020) 106877}, [\href{https://arxiv.org/abs/1901.07808}{{\ttfamily 1901.07808}}].

\bibitem{Lee:2013mka}
R.~N. Lee, \emph{{LiteRed 1.4: a powerful tool for reduction of multiloop integrals}}, \href{http://dx.doi.org/10.1088/1742-6596/523/1/012059}{\emph{J. Phys. Conf. Ser.} {\bfseries 523} (2014) 012059}, [\href{https://arxiv.org/abs/1310.1145}{{\ttfamily 1310.1145}}].

\bibitem{Bruser:2018rad}
R.~Br{\"u}ser, Z.~L. Liu and M.~Stahlhofen, \emph{{Three-Loop Quark Jet Function}}, \href{http://dx.doi.org/10.1103/PhysRevLett.121.072003}{\emph{Phys. Rev. Lett.} {\bfseries 121} (2018) 072003}, [\href{https://arxiv.org/abs/1804.09722}{{\ttfamily 1804.09722}}].

\bibitem{vonManteuffel:2014qoa}
A.~von Manteuffel, E.~Panzer and R.~M. Schabinger, \emph{{A quasi-finite basis for multi-loop Feynman integrals}}, \href{http://dx.doi.org/10.1007/JHEP02(2015)120}{\emph{JHEP} {\bfseries 02} (2015) 120}, [\href{https://arxiv.org/abs/1411.7392}{{\ttfamily 1411.7392}}].

\bibitem{Panzer:2014caa}
E.~Panzer, \emph{{Algorithms for the symbolic integration of hyperlogarithms with applications to Feynman integrals}}, \href{http://dx.doi.org/10.1016/j.cpc.2014.10.019}{\emph{Comput. Phys. Commun.} {\bfseries 188} (2015) 148--166}, [\href{https://arxiv.org/abs/1403.3385}{{\ttfamily 1403.3385}}].

\bibitem{Korchemsky:1987wg}
G.~P. Korchemsky and A.~V. Radyushkin, \emph{{Renormalization of the Wilson Loops Beyond the Leading Order}}, \href{http://dx.doi.org/10.1016/0550-3213(87)90277-X}{\emph{Nucl. Phys. B} {\bfseries 283} (1987) 342--364}.

\bibitem{Moch:2004pa}
S.~Moch, J.~A.~M. Vermaseren and A.~Vogt, \emph{{The three-loop splitting functions in {QCD}: The non-singlet case}}, \href{http://dx.doi.org/10.1016/j.nuclphysb.2004.03.030}{\emph{Nucl. Phys.} {\bfseries B688} (2004) 101--134}, [\href{https://arxiv.org/abs/hep-ph/0403192}{{\ttfamily hep-ph/0403192}}].

\bibitem{Hoang:2015vua}
A.~H. Hoang, A.~Pathak, P.~Pietrulewicz and I.~W. Stewart, \emph{{Hard Matching for Boosted Tops at Two Loops}}, \href{http://dx.doi.org/10.1007/JHEP12(2015)059}{\emph{JHEP} {\bfseries 12} (2015) 059}, [\href{https://arxiv.org/abs/1508.04137}{{\ttfamily 1508.04137}}].

\bibitem{Becher:2008cf}
T.~Becher and M.~D. Schwartz, \emph{{A Precise determination of $\alpha_s$ from LEP thrust data using effective field theory}}, \href{http://dx.doi.org/10.1088/1126-6708/2008/07/034}{\emph{JHEP} {\bfseries 07} (2008) 034}, [\href{https://arxiv.org/abs/0803.0342}{{\ttfamily 0803.0342}}].

\bibitem{Heinrich:2023til}
G.~Heinrich, S.~P. Jones, M.~Kerner, V.~Magerya, A.~Olsson and J.~Schlenk, \emph{{Numerical scattering amplitudes with pySecDec}}, \href{http://dx.doi.org/10.1016/j.cpc.2023.108956}{\emph{Comput. Phys. Commun.} {\bfseries 295} (2024) 108956}, [\href{https://arxiv.org/abs/2305.19768}{{\ttfamily 2305.19768}}].

\bibitem{Smirnov:2021rhf}
A.~V. Smirnov, N.~D. Shapurov and L.~I. Vysotsky, \emph{{FIESTA5: Numerical high-performance Feynman integral evaluation}}, \href{http://dx.doi.org/10.1016/j.cpc.2022.108386}{\emph{Comput. Phys. Commun.} {\bfseries 277} (2022) 108386}, [\href{https://arxiv.org/abs/2110.11660}{{\ttfamily 2110.11660}}].

\end{thebibliography}\endgroup
\bibliographystyle{JHEP}

\end{document}